\documentclass[preprint,pre,showpacs]{revtex4}
\usepackage{amscd}
\usepackage{amsfonts}
\usepackage{amsmath}
\usepackage{amssymb}
\usepackage{undertilde}
\usepackage{color}
\usepackage{multirow}
\usepackage{graphicx}
\usepackage{graphics}
\usepackage{wrapfig}
\usepackage[T1]{fontenc}
\usepackage[latin1]{inputenc}
\usepackage{gensymb}
\def\>{\rangle}
\def\<{\langle}
\def\n{\nonumber}

\def\sc{\scriptsize}
\newcommand{\id}{\openone}
\begin{document}
\title{R\'{e}nyi-$\alpha$ entropies of quantum states in closed form:\\Gaussian states and a class of non-Gaussian states}
\author{Ilki Kim}
\email{hannibal.ikim@gmail.com} \affiliation{Joint School of
Nanoscience and Nanoengineering, North Carolina A$\&$T State
University, Greensboro, NC 27411}
\date{\today}
\begin{abstract}
In this work, we study the R\'{e}nyi-$\alpha$ entropies
$S_{\alpha}(\hat{\rho}) = (1 -
\alpha)^{-1}\,\ln\{\mbox{Tr}(\hat{\rho}^{\alpha})\}$ of quantum
states for $N$ bosons in the phase-space representation. With the
help of the Bopp rule, we derive the entropies of Gaussian states in
closed form for positive integers $\alpha = 2,3,4,\cdots$ and then,
with the help of the analytic continuation, acquire the closed form
also for real-values of $\alpha
> 0$. The quantity $S_2(\hat{\rho})$, primarily studied in the
literature, will then be a special case of our finding. Subsequently
we acquire the R\'{e}nyi-$\alpha$ entropies, with positive integers
$\alpha$, in closed form also for a specific class of the
non-Gaussian states (mixed states) for $N$ bosons, which may be
regarded as a generalization of the eigenstates $|n\>$ (pure states)
of a single harmonic oscillator with $n \geq 1$, in which the Wigner
functions have negative values indeed. Due to the fact that the
dynamics of a system consisting of $N$ oscillators is Gaussian, our
result will contribute to a systematic study of the
R\'{e}nyi-$\alpha$ entropy dynamics when the current form of a
non-Gaussian state is initially prepared.
\end{abstract}
\maketitle
%
\section{Introduction}\label{sec:introduction}
The R\'{e}nyi-$\alpha$ entropy defined as $S_{\alpha}(\hat{\rho}) =
(1 - \alpha)^{-1}\,\ln\{\mbox{Tr}(\hat{\rho}^{\alpha})\}$ where
$\alpha
> 0$ is considered a generalization of the von-Neumann entropy $S_1(\hat{\rho})$
\cite{WEH78}. Its properties have recently been studied, e.g., in a
generalized formulation of quantum thermodynamics, which is built
upon the maximum entropy principle applied to
$S_{\alpha}(\hat{\rho})$ \cite{MIS15}, as well as in the discussion
of its time derivative under the Lindblad dynamics, the result of
which may be useful for exploring the dynamics of quantum
entanglement in the Markovian regime \cite{ABE16}. However, their
explicit expressions for $\alpha \ne 2$ have not been investigated
extensively, even for relatively simple forms of states
$\hat{\rho}$, such as the Gaussian states for $N$ bosons. In fact,
only the entropy $S_2(\hat{\rho}) = -\ln\{\mbox{Tr}(\hat{\rho}^2)\}$
has been the primary quantity for investigation thus far, where the
purity measure $\mbox{Tr}(\hat{\rho}^2)$ is the first moment of
probability $\overline{p} = \sum_j\,(p_j)^2$ with the eigenvalues
$p_j$'s of $\hat{\rho}$, whereas, e.g., the second moment
$\overline{p^2} = \mbox{Tr}(\hat{\rho}^3)$ is needed for the entropy
$S_3(\hat{\rho})$.

A Gaussian state is defined as a quantum state, the Wigner function
of which is Gaussian, such as the canonical thermal equilibrium
state of a single harmonic oscillator (including its ground state at
$T = 0$), the coherent state and the squeezed state, etc.
\cite{BRA05,FER05,WOL06,WED12,OLI12,ADE14}. The Gaussian states have
recently attracted considerable interest as the need for a better
theoretical understanding increases in response to the novel
experimental manipulation of such states in quantum optics, in
particular for the quantum information processing with continuous
variables. As is well-known, the statistical behaviors of an
$N$-mode Gaussian state are fully characterized by its covariance
matrix [cf. (\ref{eq:gaussian-state1-1})], which can yield an
evaluation of $S_2(\hat{\rho})$ straightforwardly. Besides, the
so-called Wigner entropy defined as $S_{\scriptscriptstyle
W}(\hat{\rho}) = -\int dq dp\; W_{\rho}(q,p)\, \ln\{W_{\rho}(q,p)\}$
has been in investigation for the Gaussian states in \cite{PAT17},
where the Wigner function is explicitly given by
\cite{WIG32,HIL84,LEE95,SCH01,ZAC05}
\begin{equation}\label{eq:bopp_0-1}
    W_{\rho}(q,p)\, =\, \frac{1}{2\pi\hbar} \int_{-\infty}^{\infty} d{\xi}\,
    \exp\left(-\frac{i}{\hbar} p\,\xi\right)\, \left\<q + \frac{\xi}{2}\right|\hat{\rho}\left|q - \frac{\xi}{2}\right\>
\end{equation}
for a single mode, for the simplicity of notation, as well as the
Weyl-Wigner $c$-number representation of the operator $\hat{A}$
given by
\begin{equation}
    A(q,p)\, =\, \int_{-\infty}^{\infty} d{\xi}\,
    \exp\left(-\frac{i}{\hbar} p\,\xi\right)\, \left\<q + \frac{\xi}{2}\right|\hat{A}\left|q -
    \frac{\xi}{2}\right\>\,,
\end{equation}
together giving rise to the expectation value
\begin{eqnarray}\label{eq:bopp_3}
    \<\hat{A}\>\, =\, \int dq \int dp\; W_{\rho}(q,p)\; A(q,p)\,.
\end{eqnarray}
If the operator $\hat{A} = \hat{\rho}$, then its expectation value
is nothing else than the purity measure $\mbox{Tr}(\hat{\rho}^2) =
\int dq dp\, W_{\rho}(q,p)\, A(q,p) = (2\pi\hbar) \int dq dp\;
\{W_{\rho}(q,p)\}^2$, which is the case of the R\'{e}nyi parameter
$\alpha = 2$. Interestingly, the entropy $S_2(\hat{\rho})$ for
$N$-mode Gaussian states has been shown to coincide with
$S_{\scriptscriptstyle W}(\hat{\rho})$ up to a constant
\cite{ADE12}. However, the moments $\overline{p^{\alpha-1}} =
\mbox{Tr}(\hat{\rho}^{\alpha})$ and the resulting entropies in
closed form where $\alpha > 2$ have still been unknown even for the
Gaussian states.

Therefore it will be interesting to study the R\'{e}nyi-$\alpha$
entropies in arbitrary orders for arbitrary quantum states
explicitly in the phase-space representation [cf.
(\ref{eq:bopp_0-1})-(\ref{eq:bopp_3})], and then exactly evaluate
them for some specific states such as $N$-mode Gaussian states, as
well as a certain class of non-Gaussian states, where the Wigner
function can possess negative values indeed and so the Wigner
entropy $S_{\scriptscriptstyle W}(\hat{\rho})$ is not directly
well-defined. In fact, the full knowledge of density matrix
($p_j$'s) and its diagonalization is practically hardly possible to
achieve for $N$-mode generic cases. Therefore, the phase-space
representation may also be favorable for studying the moments
$\mbox{Tr}(\hat{\rho}^{\alpha})$, with no need to diagonalize the
density matrix directly. Moreover, this phase-space approach will be
useful for studying systematically the quantum-classical transition
behaviors of the entropies. In fact, it is known that all
R\'{e}nyi-$\alpha$ entropies tend asymptotically to the von-Neumann
one in the classical limit (e.g., \cite{LIN13,BRA15,PAT17}).

The general layout of this paper is as follows: In Sec.
\ref{sec:renyi} we provide a generic framework for the moments of
density operator in the phase-space representation for arbitrary
quantum states. In Sec. \ref{sec:gaussian} we apply this rigorous
framework to $N$-mode Gaussian states and derive the
R\'{e}nyi-$\alpha$ entropies in closed form, as well as generalize
the results available in the literature. In Sec.
\ref{sec:non-gaussian} the same discussion will take place for a
class of non-Gaussian states. Finally we give the concluding remarks
of this paper in Sec. \ref{sec:conclusion}.

\section{Higher-Order Moments of Density Operator in Phase-Space Representation}\label{sec:renyi}
%
We begin with the case of the R\'{e}nyi parameter $\alpha = 3$, in
which $\mbox{Tr}(\hat{\rho}^3) = \mbox{Tr}(\hat{\rho}\,\hat{\rho}^2)
= \overline{p^2}$ is in consideration. To do so, we apply the Bopp
rule for the Weyl-Wigner representation of the operator product
$\hat{B}_1\hat{B}_2$, explicitly given by \cite{SCH01}
\begin{equation}\label{eq:bopp1}
    (B_1B_2)(q,p) = B_1\left(q-\frac{\hbar}{2i}\frac{\partial}{\partial p},
    p+\frac{\hbar}{2i}\frac{\partial}{\partial
    q}\right)\,B_2(q,p)
\end{equation}
(for a single mode), in which $\hat{B}_1 = \hat{\rho}$ and
$\hat{B}_2 = \hat{\rho}$ for our purpose. Then we have, with
$\hat{A} = \hat{B}_1\hat{B}_2$, the second moment of probability
\begin{eqnarray}\label{eq:bopp_4}
    \overline{p^2} &=& \int dq dp\, W_{\rho}(q,p)\, A(q,p)\n\\
    &=& (2\pi\hbar)^2 \int dq dp\, W_{\rho}(q,p)\, W_{\rho}\left(q-\frac{\hbar}{2i}\frac{\partial}{\partial p},
    p+\frac{\hbar}{2i}\frac{\partial}{\partial q}\right)\,
    W_{\rho}(q,p)\,.
\end{eqnarray}
With the help of the Taylor expansion given by $W_{\rho}(q+h_q,
p+h_p) = \sum_{k=0}^{\infty} (1/k!) (h_q\,\partial_q +
h_p\,\partial_p)^k\,W_{\rho}(q,p)$ with $h_q =
-(\hbar/2i)\,\partial_p$ and $h_p = (\hbar/2i)\,\partial_x$, we can
obtain, on the right-hand side of (\ref{eq:bopp_4}),
\begin{equation}\label{eq:mit-math-ency}
    W_{\rho}\left(q-\frac{\hbar}{2i}\frac{\partial}{\partial p}, p+\frac{\hbar}{2i}\frac{\partial}{\partial q}\right)\,
    W_{\rho}(q,p)
    = \sum_{k=0} \frac{1}{k!} \left(\frac{i\hbar}{2}\right)^k \left(\partial_{1q}\,\partial_{2p} -
    \partial_{1p}\,\partial_{2q}\right)^k\, W_1(q,p)\,
    W_2(q,p)\,,
\end{equation}
where the operators $\partial_1$ and $\partial_2$ affect $W_1 =
W_{\rho}$ and $W_2 = W_{\rho}$ alone, respectively. Due to the
symmetry between $W_1$ and $W_2$, all terms with $k$ odd can be
shown to vanish indeed. It is also easy to see that in the classical
limit of $\hbar \to 0$, only the term of $k=0$ is non-vanishing, and
so Eq. (\ref{eq:mit-math-ency}) will reduce to its classical
counterpart $\{W_{\mbox{\sc cl}}(q,p)\}^2$.

Now we generalize this single-mode expression into that of $N$
modes. Let $\vec{q} = (q_1, q_2, \cdots, q_{\scriptscriptstyle
N})^{\scriptscriptstyle T}$ and $\vec{p} = (p_1, p_2, \cdots,
p_{\scriptscriptstyle N})^{\scriptscriptstyle T}$. Then, Eq.
(\ref{eq:mit-math-ency}) will easily be transformed into
\begin{eqnarray}\label{eq:mit-math-ency1}
    && W_{\rho}\left(\sum_{n=1}^N \left(q_n-\frac{\hbar}{2i}\frac{\partial}{\partial p_n}\right),
    \sum_{n=1}^N \left(p_n+\frac{\hbar}{2i}\frac{\partial}{\partial q_n}\right)\right)\cdot
    W_{\rho}(\vec{q},\vec{p})\n\\
    &=& \sum_{k=0} \frac{1}{k!} \left(\frac{i\hbar}{2}\right)^k \sum_{n=1}^N
    \left(\partial_{1q_n} \partial_{2p_n} - \partial_{1p_n} \partial_{2q_n}\right)^k\, W_1(\vec{q},\vec{p})\, W_2(\vec{q},\vec{p})\,.
\end{eqnarray}
We substitute into (\ref{eq:mit-math-ency1}) the expression
\begin{equation}\label{eq:fourier-transform-1}
    W_{\rho}(\vec{q},\vec{p}) = \int \frac{d^{\scriptscriptstyle N}\vec{{\mathfrak q}}\,d^{\scriptscriptstyle
    N}\vec{{\mathfrak p}}}{(2\pi\hbar)^{\scriptscriptstyle N}}\,
    \widetilde{W}_{\rho}(\vec{{\mathfrak q}},\vec{{\mathfrak p}})\,
    \exp\left\{-\frac{i}{\hbar}\sum_n^N (q_n {\mathfrak p_n} + p_n {\mathfrak q_n})\right\}
\end{equation}
where the symbol $\widetilde{W}_{\rho}(\vec{{\mathfrak
q}},\vec{{\mathfrak p}})$ denotes the Fourier transform of
$W_{\rho}(\vec{q},\vec{p})$. After making some algebraic
manipulations, we can finally transform (\ref{eq:mit-math-ency1})
into the compact form
\begin{equation}\label{eq:fourier-transform-2}
    \int \frac{d^{2{\scriptscriptstyle N}}\vec{x}_1}{(2\pi\hbar)^{\scriptscriptstyle N}}
    \frac{d^{2{\scriptscriptstyle N}}\vec{x}_2}{(2\pi\hbar)^{\scriptscriptstyle
    N}}\,\widetilde{W}_{\rho}(\vec{x}_1)\, \widetilde{W}_{\rho}(\vec{x}_2)\,
    \exp\left\{-\frac{i}{\hbar}\,(\vec{x})^{\scriptscriptstyle T}\,\hat{\Lambda}\,(\vec{x}_1 +
    \vec{x}_2)\right\}\,
    \exp\left\{\frac{i}{2\hbar} (\vec{x}_1)^{\scriptscriptstyle T}\,\hat{\Omega}\,
    \vec{x}_2\right\}\,,
\end{equation}
in which the vector $\vec{x} = (q_1, p_1, \cdots,
q_{\scriptscriptstyle N}, p_{\scriptscriptstyle
N})^{\scriptscriptstyle T} \in {\mathbb R}^{2{\scriptscriptstyle
N}}$,
and $\hat{\Lambda} = \oplus_{n=1}^{\scriptscriptstyle N} \bigl(\begin{smallmatrix} 0&1 \\
1&0 \end{smallmatrix} \bigr)$, as well as $\hat{\Omega} =
\oplus_{n=1}^{\scriptscriptstyle N} \bigl(\begin{smallmatrix} 0&1 \\
-1&0
\end{smallmatrix} \bigr)$; for a single mode,
we have the reduced expressions here, $(\vec{x})^{\scriptscriptstyle
T}\,\hat{\Lambda}\,(\vec{x}_1 + \vec{x}_2) \to q\,(p_1 + p_2) +
p\,(q_1 + q_2)$ and $(\vec{x}_1)^{\scriptscriptstyle
T}\,\hat{\Omega}\,\vec{x}_2 \to q_1\,p_2 - p_1\,q_2$. It is also
easy to note that this integral form is real-valued, by exchanging
the variables ($\vec{x}_1 \leftrightarrow \vec{x}_2$) with the
relation $(\vec{x}_1)^{\scriptscriptstyle
T}\,\hat{\Omega}\,\vec{x}_2 = -(\vec{x}_2)^{\scriptscriptstyle
T}\,\hat{\Omega}\,\vec{x}_1$. Then the second moment is given by
[cf. (\ref{eq:bopp_4})-(\ref{eq:mit-math-ency1}) and
(\ref{eq:fourier-transform-2})]
\begin{equation}\label{eq:psquared-bar_0}
    \overline{p^2} = (2\pi\hbar)^{\scriptscriptstyle N} \int d^{2{\scriptscriptstyle
    N}}\vec{x}\;
    W_{\rho}(\vec{x})\, W_{\rho^2}(\vec{x})\,,
\end{equation}
where $W_{\rho^2}(\vec{x})$ denotes the integral form
(\ref{eq:fourier-transform-2}) multiplied by
$(2\pi\hbar)^{\scriptscriptstyle N}$.

Along the lines similar to the case of $\alpha = 3$, we can study
the next case of $\alpha =4$. By applying the Bopp rule
(\ref{eq:bopp1}) for $A(q,p) = (2\pi\hbar)^3\,(B_1 B_2 B_3)(q,p)$
with $\hat{B}_1 = \hat{B}_2 = \hat{B}_3 = \hat{\rho}$, we can have
the third moment of probability
\begin{eqnarray}\label{eq:bopp_4-2}
    && \overline{p^3}\, =\, \int dq dp\, W_{\rho}(q,p)\, A(q,p)\\
    &=& (2\pi\hbar)^3 \int dq dp\, W_{\rho}(q,p)\, W_{\rho}\left(q-\frac{\hbar}{2i}\frac{\partial}{\partial p},
    p+\frac{\hbar}{2i}\frac{\partial}{\partial q}\right)\, W_{\rho}\left(q-\frac{\hbar}{2i}\frac{\partial}{\partial p},
    p+\frac{\hbar}{2i}\frac{\partial}{\partial q}\right)\,
    W_{\rho}(q,p)\n
\end{eqnarray}
for a single mode. Following the steps provided in
(\ref{eq:mit-math-ency})-(\ref{eq:fourier-transform-2}) for the
second moment, it is straightforward to derive an expression of the
third moment for $N$ modes, which will exactly be the counterpart to
(\ref{eq:psquared-bar_0}). We easily observe that the same
techniques will be employed also for $\alpha = 5,6,7, \cdots$. In
fact, we can finally derive an expression of the $j$th moment, with
$j = \alpha-1$, for $N$ modes
\begin{equation}\label{eq:bopp_4-3}
    \overline{p^j} = \mbox{Tr}(\hat{\rho}^{j+1}) = (2\pi\hbar)^{\scriptscriptstyle N} \int d^{2{\scriptscriptstyle N}}\vec{x}\, W_{\rho}(\vec{x})\,
    W_{\rho^{j}}(\vec{x})\,,
\end{equation}
which is valid for an arbitrary state $\hat{\rho}$ with $j =
1,2,3,\cdots$ [cf. (\ref{eq:psquared-bar_0})]. Here, a factor of the
integrand
\begin{eqnarray}\label{eq:bopp_4-3-1}
    \hspace*{-0.8cm}&& W_{\rho^j}(\vec{x})\, =\, (2\pi\hbar)^{(j-1){\scriptscriptstyle N}}\,
    \left[W_{\rho}\left(\sum_{n=1}^N \left(q_n-\frac{\hbar}{2i}\frac{\partial}{\partial p_n}\right),
    \sum_{n=1}^N \left(p_n+\frac{\hbar}{2i}\frac{\partial}{\partial q_n}\right)\right)\right]^{j-1}\, W_{\rho}(\vec{x})\\
    \hspace*{-0.8cm}&=& \int \frac{d^{2{\scriptscriptstyle N}}\vec{x}_1 \cdots
    d^{2{\scriptscriptstyle N}}\vec{x}_j}{(2\pi\hbar)^{\scriptscriptstyle N}} \left\{\prod_{\nu=1}^j
    \widetilde{W}_{\rho}(\vec{x}_{\nu})\right\}\,
    \exp\left\{-\frac{i}{\hbar}\,(\vec{x})^{\scriptscriptstyle T}\,\hat{\Lambda}\,\sum_{\nu}^j
    \vec{x}_{\nu}\right\}\,
    \exp\left\{\frac{i}{2\hbar}\sum_{\nu < \mu}^j (\vec{x}_{\nu})^{\scriptscriptstyle
    T}\,\hat{\Omega}\,\vec{x}_{\mu}\right\}\n
\end{eqnarray}
[cf. (\ref{eq:fourier-transform-2}) for $j=2$\,]. Consequently, we
can explicitly discuss the R\'{e}nyi-$\alpha$ entropies
$S_{\alpha}({\hat{\rho}})$ where $\alpha = j+1 = 2,3,4,\cdots$. In
the next sections, we will evaluate those moments of probability and
the corresponding entropies in closed form for some specific states.

For comparison, we briefly consider two additional entropies now.
The first one is the von-Neumann entropy $S_{\mbox{\sc
vN}}(\hat{\rho})$. It is easy to rewrite this entropy as
\begin{eqnarray}\label{eq:det1-5}
    S_1(\hat{\rho}) &=& -\sum_{\nu} p_{\nu}\, \ln\{\overline{p}\,(p_{\nu}/\overline{p})\}\, =\,
    -\ln(\overline{p}) +
    \sum_{\nu} p_{\nu} \sum_{{\mu}=2}\,\frac{1}{\mu}\, \{1 -
    p_{\nu}\,(\overline{p})^{-1}\}^{\mu}\n\\
    &=& S_2(\hat{\rho}) + \sum_{\mu=2}
    \frac{(-1)^{\mu}\,{\mathcal M}_{\mu}}{\mu\, (\overline{p})^{\mu}}\,,
\end{eqnarray}
where the central moments ${\mathcal M}_{\mu} := \overline{(p -
\overline{p})^{\mu}} = \mbox{Tr}(\hat{\rho}\,\{\hat{\rho} -
\mbox{Tr}(\hat{\rho}^2)\}^{\mu})$. Therefore, the difference between
$S_1(\hat{\rho})$ and $S_2(\hat{\rho})$ is explicitly given by that
sum of all higher-order moments, each of which can be evaluated with
the help of the Bopp rule, as shown. The second entropy is the
Wigner entropy (for an $N$-mode state), explicitly given by
\begin{equation}\label{eq:wigner-entropy-1}
    S_{\scriptscriptstyle W}(\hat{\rho}) = -\int d^{2{\scriptscriptstyle N}}\vec{x}\; W_{\rho}(\vec{x})\;
    \ln\{\hbar^{{\scriptscriptstyle N}}\,W_{\rho}(\vec{x})\}\,,
\end{equation}
which is well-defined so long as $W_{\rho}(\vec{x}) \geq 0$ over the
phase space. This also can be expanded as
\begin{equation}\label{eq:wigner-entropy-expanded-1}
    S_{\scriptscriptstyle W}(\hat{\rho}) = \sum_{\nu=1} \frac{(-1)^{\nu}}{\nu}
    \int d^{2{\scriptscriptstyle N}}\vec{x}\; W_{\rho}(\vec{x})\, \{\hbar^{\scriptscriptstyle N}\,W_{\rho}(\vec{x}) - 1\}^{\nu}\,.
\end{equation}
As seen, the Bopp rule was not at all employed here for each $\nu$
such that $\int \{W_{\rho}(\vec{x})\}^{\nu} \not\propto
\mbox{Tr}(\hat{\rho}^{\nu})$ where $\nu \geq 3$. Therefore, the
Wigner entropy, albeit well-defined, cannot directly be related to
the von-Neumann entropy.

\section{R\'{e}nyi Entropies of $N$-Mode Gaussian States}\label{sec:gaussian}
%
We first consider the case of an $N$-mode Gaussian state, which is
explicitly given by \cite{OLI12,ADE14}
\begin{equation}\label{eq:gaussian-state1-1}
    W_{\rho}(\vec{x}) = \frac{\exp\{-(\vec{x}-\vec{d})^{\scriptscriptstyle T}\,(\hat{\sigma})^{-1}\,
    (\vec{x}-\vec{d})\}}{{\pi^{\scriptscriptstyle
    N}\,\{\det(\hat{\sigma})\}}^{1/2}}\,,
\end{equation}
in which the vector $\vec{x} \in {\mathbb R}^{2{\scriptscriptstyle
N}}$ as defined above, and the first moments $d_j := \overline{x}_j
= \mbox{Tr}(\hat{\rho}\,\hat{{\mathcal X}}_j)$ where the vector of
operators $\vec{{\mathcal X}} = (\hat{Q}_1, \hat{P}_1, \cdots,
\hat{Q}_{\scriptscriptstyle N}, \hat{P}_{\scriptscriptstyle
N})^{\scriptscriptstyle T}$, as well as the $2N$-by-$2N$ matrix
$\hat{\sigma} := 2\,\utilde{\hat{\sigma}}$ where the covariance
matrix $\utilde{\hat{\sigma}}$ denotes the corresponding second
moments, given by $\utilde{\sigma}_{jk} =
\mbox{Tr}[\hat{\rho}\,(\hat{{\mathcal X}}_j\,\hat{{\mathcal X}}_k +
\hat{{\mathcal X}}_k\,\hat{{\mathcal X}}_j)/2] - d_j\,d_k$. Then,
$\mbox{Tr}(\hat{\rho}^2) = \hbar^{\scriptscriptstyle
N}\,\{\mbox{det}(\sigma)\}^{-1/2}$. As is well-known, the Fourier
transform of (\ref{eq:gaussian-state1-1}) is Gaussian, too, and
explicitly given by
\begin{equation}\label{eq:gaussian-state1-1-1}
    \widetilde{W}_{\rho}(\vec{x}) =
    \frac{\exp\{-(4\hbar^2)^{-1}\,(\vec{x}-\vec{d})^{\scriptscriptstyle
    T}\,\hat{\Lambda}\,\hat{\sigma}\,\hat{\Lambda}\,(\vec{x}-\vec{d})\}}{(2\pi\hbar)^{{\scriptscriptstyle
    N}}}\,,
\end{equation}
which can be acquired by applying the $L$-dimensional Gaussian
integral for $L = 2N$ \cite{ZEE03}
\begin{equation}\label{eq:gaussian-integral_zee}
    \int_{-\infty}^{\infty} dy_1 \cdots \int_{-\infty}^{\infty}
    dy_{\scriptscriptstyle L}\, \exp\left\{-\frac{1}{2}\,
    (\vec{y})^{\scriptscriptstyle T}\,\hat{\Upsilon}\, \vec{y} +
    (\vec{\upsilon})^{{\scriptscriptstyle T}}\,\vec{y}\right\} =
    \left\{\frac{(2\pi)^L}{\mbox{det}(\hat{\Upsilon})}\right\}^{1/2} \exp\left\{\frac{1}{2}\,
    (\vec{\upsilon})^{\scriptscriptstyle T}\,\hat{\Upsilon}^{-1}\,\vec{\upsilon}\right\}
\end{equation}
with $\vec{y}, \vec{\upsilon} \in {\mathbb R}^{{\scriptscriptstyle
L}}$. Here the symbol $\hat{\Upsilon}$ denotes a symmetric
$L$-by-$L$ matrix. Since the first moments $d_j$'s give rise to the
displacement of Wigner function (\ref{eq:gaussian-state1-1}) but do
not change its shape, we will set $\vec{d} = 0$ from now on, without
loss of generality for our discussion of the probability moments and
R{\'{e}nyi-$\alpha$ entropies which remain unchanged with respect to
this displacement.

Now we substitute
(\ref{eq:gaussian-state1-1})-(\ref{eq:gaussian-state1-1-1}) into the
framework (\ref{eq:bopp_4-3})-(\ref{eq:bopp_4-3-1}) and then apply
(\ref{eq:gaussian-integral_zee}) for $L= 2N(j+1)$ with
$\vec{\upsilon} = 0$ and $dy_1 \cdots dy_{{\scriptscriptstyle L}} =
d^{2{\scriptscriptstyle N}}\vec{x} \prod_{\nu=1}^j
d^{2{\scriptscriptstyle N}}\vec{x}_{\nu}$ in order to acquire the
probability moments $\overline{p^j}$ in closed form. This will
easily result in
\begin{equation}\label{eq:jth-moment-1}
    \overline{p^j} = \mbox{Tr}(\hat{\rho}^{j+1}) = (2\hbar)^{-j{\scriptscriptstyle N}}\, \{\mbox{det}(\sigma)\}^{-1/2}\,
    \{\mbox{det}(\hat{D}_{j+1})\}^{-1/2}\,,
\end{equation}
where $\hat{\Upsilon} \to \hat{D}_{j+1}$ in
(\ref{eq:gaussian-integral_zee}) can be expressed as a
($j+1$)-by-($j+1$) block symmetric matrix, each block of which is a
$2N$-by-$2N$ matrix; for the simplest case of ($N=1, j=1$), we
explicitly have the 4-by-4 symmetric matrix
\begin{equation}\label{eq:D_2_N=1}
    \hat{D}_2 = \left(\begin{array}{cccc}
              {\displaystyle \frac{\sigma_{22}}{4\hbar^2}} & {\displaystyle \frac{\sigma_{12}}{4\hbar^2}} & 0 & {\displaystyle \frac{i}{2\hbar}}\\[2ex]
              {\displaystyle \frac{\sigma_{12}}{4\hbar^2}} & {\displaystyle \frac{\sigma_{11}}{4\hbar^2}} & {\displaystyle \frac{i}{2\hbar}} & 0\\[2ex]
              0 & {\displaystyle \frac{i}{2\hbar}} & {\displaystyle \frac{\sigma_{22}}{\mbox{det}(\hat{\sigma})}} & {\displaystyle \frac{-\sigma_{12}}{\mbox{det}(\hat{\sigma})}}\\[2ex]
              {\displaystyle \frac{i}{2\hbar}} & 0 & {\displaystyle \frac{-\sigma_{12}}{\mbox{det}(\hat{\sigma})}} & {\displaystyle
              \frac{\sigma_{11}}{\mbox{det}(\hat{\sigma})}}
    \end{array}\right)\,,
\end{equation}
where $\mbox{det}(\hat{\sigma}) = \sigma_{11}\,\sigma_{22} -
(\sigma_{12})^2$. The matrices $\hat{D}_{j+1}$'s for $N$ modes are
formally provided in Tab. \ref{tab:table1}.
\begin{table}[ht]
\caption{The symmetric $2N(j+1)$-by-$2N(j+1)$ matrix
$\hat{D}_{j+1}$, used for an evaluation of $\overline{p^j}$ in
(\ref{eq:jth-moment-1}), is now expressed as a ($j+1$)-by-($j+1$)
block matrix, where $j = 1, 2, 3, 4, 5, 6$. Here $\vec{x} \in
{\mathbb R}^{2{\scriptscriptstyle N}}$ and $\vec{x}_j \in {\mathbb
R}^{2{\scriptscriptstyle N}}$. Consequently, $\hat{D}_7$ is a 7-by-7
block matrix covering ($\vec{x}, \vec{x}_1, \cdots, \vec{x}_6$);
$\hat{D}_6$ is a 6-by-6 block matrix covering ($\vec{x}, \vec{x}_1,
\cdots, \vec{x}_5$); $\cdots$; $\hat{D}_2$ is a 2-by-2 block matrix
covering ($\vec{x}, \vec{x}_1$). Besides, $\hat{D}_1$ is a 1-by-1
block matrix covering $\vec{x}$ only. In fact, $\hat{D}_1 =
(\sigma)^{-1}$ as seen, and $\mbox{Tr}(\hat{\rho}) = 1$ in
(\ref{eq:jth-moment-1}), thus representing the trivial case. The
explicit expression of $\hat{D}_{j+1}$ for $j = 7,8,9,\cdots$ will
easily follow in the same way.}
\begin{center}
    $\hat{D}_7 =$
    \begin{tabular}{||c|c|c|c|c|c|c||c}
        $\vec{x}_6$ & $\vec{x}_5$ & $\vec{x}_4$ & $\vec{x}_3$ & $\vec{x}_2$ & $\vec{x}_1$ & $\vec{x}$ &\\
        \hline\hline
        $\hat{\Lambda}\hat{\sigma}\hat{\Lambda}/(4\hbar^2)$ & $\hat{\Omega}/(4 i\hbar)$ & $\hat{\Omega}/(4
        i\hbar)$
        & $\hat{\Omega}/(4 i\hbar)$ & $\hat{\Omega}/(4 i\hbar)$ & $\hat{\Omega}/(4 i\hbar)$ & $i \hat{\Lambda}/(2 \hbar)$ & $\vec{x}_6$\\ \hline
        $i \hat{\Omega}/(4 \hbar)$ & $\hat{\Lambda}\hat{\sigma}\hat{\Lambda}/(4\hbar^2)$ & $\hat{\Omega}/(4 i\hbar)$ &
        $\hat{\Omega}/(4 i\hbar)$
        & $\hat{\Omega}/(4 i\hbar)$ & $\hat{\Omega}/(4 i\hbar)$ & $i \hat{\Lambda}/(2 \hbar)$ & $\vec{x}_5$\\ \hline
        $i \hat{\Omega}/(4 \hbar)$ & $i \hat{\Omega}/(4 \hbar)$ & $\hat{\Lambda}\hat{\sigma}\hat{\Lambda}/(4\hbar^2)$ & $\hat{\Omega}/(4 i\hbar)$ &
        $\hat{\Omega}/(4 i\hbar)$
        & $\hat{\Omega}/(4 i\hbar)$ & $i \hat{\Lambda}/(2 \hbar)$ & $\vec{x}_4$\\ \hline
        $i \hat{\Omega}/(4 \hbar)$ & $i \hat{\Omega}/(4 \hbar)$ & $i \hat{\Omega}/(4 \hbar)$ & $\hat{\Lambda}\hat{\sigma}\hat{\Lambda}/(4\hbar^2)$ &
        $\hat{\Omega}/(4 i\hbar)$ &
        $\hat{\Omega}/(4 i\hbar)$
        & $i \hat{\Lambda}/(2 \hbar)$ & $\vec{x}_3$\\ \hline
        $i \hat{\Omega}/(4 \hbar)$ & $i \hat{\Omega}/(4 \hbar)$ & $i \hat{\Omega}/(4 \hbar)$ & $i \hat{\Omega}/(4 \hbar)$ &
        $\hat{\Lambda}\hat{\sigma}\hat{\Lambda}/(4\hbar^2)$ &
        $\hat{\Omega}/(4 i\hbar)$
        & $i \hat{\Lambda}/(2 \hbar)$ & $\vec{x}_2$\\ \hline
        $i \hat{\Omega}/(4 \hbar)$ & $i \hat{\Omega}/(4 \hbar)$ & $i \hat{\Omega}/(4 \hbar)$ & $i \hat{\Omega}/(4 \hbar)$ & $i \hat{\Omega}/(4 \hbar)$ &
        $\hat{\Lambda}\hat{\sigma}\hat{\Lambda}/(4\hbar^2)$
        & $i \hat{\Lambda}/(2 \hbar)$ & $\vec{x}_1$\\ \hline
        $i \hat{\Lambda}/(2 \hbar)$ & $i \hat{\Lambda}/(2 \hbar)$ & $i \hat{\Lambda}/(2 \hbar)$ &
        $i \hat{\Lambda}/(2 \hbar)$ &
        $i \hat{\Lambda}/(2 \hbar)$ & $i \hat{\Lambda}/(2 \hbar)$ & $(\hat{\sigma})^{-1}$ & $\vec{x}$\\ \hline\hline
\end{tabular}
\end{center}
\label{tab:table1}
\end{table}
\begin{figure}[ht]
    \caption{The structure of block matrix $\hat{D}_{j+1}$, used for an evaluation of its determinant.}
\begin{center}
    $\hat{D}_{j+1}$ =
\begin{tabular}{|c|ccccccccccccc|}
    \hline
    $\hat{A}$&&&&&&$\hat{B}_j$&&&&&&&\\ \hline
    &&&&&&&&&&&&&\\
    $\hat{C}_j$&&&&&&$\hat{D}_j$&&&&&&&\\
    &&&&&&&&&&&&&\\
    \hline
\end{tabular}
\end{center}
\label{fig:fig1}
\end{figure}

Now we will evaluate $\mbox{det}(\hat{D}_{j+1})$ with the help of
the recurrence relation \cite{BER09}
\begin{equation}\label{eq:matrix-determinant-1}
    \mbox{det}(\hat{D}_{j+1}) =
    \mbox{det}(\hat{G}_j)\, \mbox{det}(\hat{D}_j) = \mbox{det}(\hat{G}_j)\, \mbox{det}(\hat{G}_{j-1}) \cdots
    \mbox{det}(\hat{G}_1)\, \{\mbox{det}(\hat{\sigma})\}^{-1},
\end{equation}
in which the generator $\mbox{det}(\hat{G}_j) = \mbox{det}\{\hat{A}
- \hat{B}_j\,(\hat{D}_j)^{-1}\,\hat{C}_j\}$ (cf. Fig.
\ref{fig:fig1}). Here $\hat{A} =
\hat{\Lambda}\hat{\sigma}\hat{\Lambda}/(4\hbar^2)$ is a 1-by-1 block
matrix, as shown in Tab. \ref{tab:table1}; $\hat{B}_j$ is a 1-by-$j$
block matrix; $\hat{C}_j$ is a $j$-by-1 block matrix. Eq.
(\ref{eq:jth-moment-1}) then reduces to a simple recursive form
\begin{equation}\label{eq:matrix-determinant-2}
    \overline{p^j} = (2\hbar)^{-j{\scriptscriptstyle N}}\, \{\mbox{det}(\hat{G}_j)\, \mbox{det}(\hat{G}_{j-1}) \cdots
    \mbox{det}(\hat{G}_1)\}^{-1/2} = (2\hbar)^{-{\scriptscriptstyle N}}\,
    \{\mbox{det}(\hat{G}_j)\}^{-1/2}\; \overline{p^{j-1}}\,,
\end{equation}
where $\overline{p^0} = 1$. Now it is easy to see that the key
ingredient is the generator $\mbox{det}(\hat{G}_j)$. After some
algebraic manipulations, every single step of which is provided in
detail in the Appendix, we can finally arrive at the closed
expression
\begin{eqnarray}\label{eq:det1-1}
    \mbox{det}(\hat{G}_j) &=& \frac{{\mathcal
    D}_{\sigma}}{(2\hbar)^{2{\scriptscriptstyle N}}} \left[\frac{1}{2} + ({\mathcal
    D}_{\sigma})^{-1/2}\,
    \left\{\left[1 -
    \left\{\frac{({\mathcal D}_{\sigma})^{1/2} - 1}{({\mathcal D}_{\sigma})^{1/2} + 1}\right\}^j\right]^{-1} -
    \frac{1}{2}\right\}\right]^2\n\\
    &=& \frac{1}{(2\hbar)^{2{\scriptscriptstyle N}}}\,
    \left\{\frac{F(j+1)}{2\,F(j)}\right\}^2\,,
\end{eqnarray}
in which the dimensionless quantities ${\mathcal D}_{\sigma} :=
(\mbox{det}\,\hat{\sigma})/\hbar^{2{\scriptscriptstyle N}}$ and
$F_{\sigma}(j) := \{({\mathcal D}_{\sigma})^{1/2} + 1\}^{j} -
\{({\mathcal D}_{\sigma})^{1/2} - 1\}^j$. Plugging this into
(\ref{eq:matrix-determinant-2}), we can then obtain the $j$th
moments of probability in closed form for an arbitrary $N$-mode
Gaussian state, given by
\begin{equation}\label{eq:det1-2}
    \overline{p^j} = 2^{j+1}\,\{F_{\sigma}(j+1)\}^{-1}\,.
\end{equation}
We easily observe that if ${\mathcal D}_{\sigma} = 1$, equivalent to
$\hat{\rho}$ denoting a pure state and thus leading to
$F_{\sigma}(j+1) = 2^{j+1}$, then $\overline{p^j} = 1$ for all
$j$'s. Some of the probability moments are explicitly provided in
Tab. {\ref{tab:table2}.
\begin{table}[ht]
\caption{Explicit expressions of the probability moments for an
arbitrary $N$-mode Gaussian state.}
\begin{center}
    \begin{tabular}{|lllll|}
    \hline
    $\overline{p^0}$ & $=$ & $\mbox{Tr}(\hat{\rho}^1)$ & $=$ & $1$\\
    \hline
    $\overline{p^1}$ & $=$ & $\mbox{Tr}(\hat{\rho}^2)$ & $=$ & $({\mathcal D}_{\sigma})^{-1/2}$\\
    \hline
    $\overline{p^2}$ & $=$ & $\mbox{Tr}(\hat{\rho}^3)$ & $=$ & $4\,(3\,{\mathcal D}_{\sigma} + 1)^{-1}$\\
    \hline
    $\overline{p^3}$ & $=$ & $\mbox{Tr}(\hat{\rho}^4)$ & $=$ &
    $2\,\{({\mathcal D}_{\sigma})^{1/2}\, ({\mathcal D}_{\sigma} + 1)\}^{-1}$\\
    \hline
    $\overline{p^4}$ & $=$ & $\mbox{Tr}(\hat{\rho}^5)$ & $=$ & $16\,\{1 + 10\,{\mathcal D}_{\sigma} + 5\,({\mathcal
    D}_{\sigma})^2\}^{-1}$\\
    \hline
    $\overline{p^5}$ & $=$ & $\mbox{Tr}(\hat{\rho}^6)$ & $=$ & $16\,\{({\mathcal D}_{\sigma})^{1/2}\, ({\mathcal D}_{\sigma} + 3) (3\,{\mathcal D}_{\sigma} +
    1)\}^{-1}$\\
    \hline
    $\overline{p^6}$ & $=$ & $\mbox{Tr}(\hat{\rho}^7)$ & $=$ & $64\,\{1 + 21\,{\mathcal D}_{\sigma} + 35\,({\mathcal
    D}_{\sigma})^2 + 7\,({\mathcal D}_{\sigma})^3\}^{-1}$\\ \hline
\end{tabular}
\end{center}
\label{tab:table2}
\end{table}
As expected, all higher-order moments are expressed in terms of the
first moment $\overline{p}$. The resulting R\'{e}nyi-$\alpha$
entropies are plotted in Fig. \ref{fig:fig2}. For comparison, it is
instructive to note that for a (classical) Gaussian distribution
$f(y)$, the moments $M_l = \overline{(y-\overline{y})^l} = 0$ for
$l$ odd while $M_l = (l-1)!!\, (M_2)^{l/2}$ for $l$ even such that
$M_4 = 3\,(M_2)^2, M_6 = 15\,(M_2)^3, \cdots$, as is well-known
\cite{PAP65}. A comment is deserved here. Alternatively, we can
derive Eq. (\ref{eq:det1-1}) for a single mode only and then
generalize this into its $N$-mode counterpart, thanks to the
Williamson decomposition \cite{WIL36,OLI12,ADE14}.

In comparison with the Williamson decomposition and for a later
purpose, we consider a simple system in a single-mode Gaussian state
now. This system is a linear oscillator coupled at an arbitrary
strength to a heat bath consisting of $N_b$ uncoupled oscillators
where $N_b \gg 1$ (the Brownian oscillator) \cite{ING98}. The {\em
total} system of $N = N_b + 1$ modes, composed of oscillator
$(\hat{Q}, \hat{P})$ and bath $\{(\hat{Q}_n, \hat{P}_n)| n = 1, 2,
\cdots, N_b\}$, is assumed to be in the canonical thermal
equilibrium state $\hat{\rho}_{\beta} \propto \exp(-\beta \hat{H})$
where $\beta = 1/(k_{{\mbox{\sc B}}} T)$ and the total-system
Hamiltonian $\hat{H}$. Then the reduced state of the coupled
oscillator only is given by \cite{WEI99}
\begin{equation}\label{eq:strong-coupling-1}
    \<q|\hat{{\mathsf R}}|q'\> = \frac{1}{\sqrt{2\pi\,\<\hat{Q}^2\>_{\beta}}}\,
    \exp\left\{-\frac{(q+q')^2}{8\,\<\hat{Q}^2\>_{\beta}} -
    \frac{\<\hat{P}^2\>_{\beta}\,(q-q')^2}{2\hbar^2}\right\}\,,
\end{equation}
in which $\<\hat{Q}^2\>_{\beta} = \mbox{Tr}(\hat{\rho}_{\beta}
\hat{Q}^2)$ and $\<\hat{P}^2\>_{\beta} =
\mbox{Tr}(\hat{\rho}_{\beta} \hat{P}^2)$. This density matrix is, in
general, {\em not} in diagonal form in the energy eigenbasis
\cite{KIM10}. In the limit of vanishingly weak coupling, this state
reduces to the canonical equilibrium state, $\hat{{\mathsf R}} \to
\hat{{\mathsf R}}_{\beta}$ indeed, as required, thus yielding
$\<\hat{Q}^2\>_{\beta} \to (2\kappa^2)^{-1}\,
\mbox{coth}(\beta\hbar\omega/2)$ and $\<\hat{P}^2\>_{\beta} \to
(\hbar^2 \kappa^2/2)\, \mbox{coth}(\beta\hbar\omega/2)$ where
$\kappa := (M \omega/\hbar)^{1/2}$, as is well-known \cite{TAN07}.

The Wigner representation of (\ref{eq:strong-coupling-1}) is then
given by a Gaussian form
\begin{equation}\label{eq:Wigner-strong-coupling-1}
    W_{\scriptscriptstyle {\mathsf R}}(q,p) =
    \frac{1}{2\pi \sqrt{\<\hat{Q}^2\>_{\beta}\,\<\hat{P}^2\>_{\beta}}}\,
    \exp\left\{-\frac{1}{2}\,\left(\frac{q^2}{\<\hat{Q}^2\>_{\beta}}
    + \frac{p^2}{\<\hat{P}^2\>_{\beta}}\right)\right\}\,,
\end{equation}
where the (2-by-2) covariance matrix is in diagonal form,
$(\hat{\utilde{\sigma}})_{jk} = \utilde{\sigma}_j\,\delta_{jk}$ with
$\utilde{\sigma}_1 = {\<\hat{Q}^2\>_{\beta}}$ and $\utilde{\sigma}_2
= {\<\hat{P}^2\>_{\beta}}$. It follows that ${\mathcal D}_{\sigma} =
(2 v)^2$ where $v =
(\utilde{\sigma}_1\,\utilde{\sigma}_2)^{1/2}/\hbar$. Only in the
limit of vanishingly weak coupling (i.e., the canonical thermal
equilibrium), we have $v = \overline{n} + (1/2)$ where the average
number of quanta $\overline{n} = \{\exp(\beta \hbar \omega) -
1\}^{-1}$, as well as the two components become identical,
$\utilde{\sigma}_1 = \utilde{\sigma}_2$ with $\hbar = 1$ and $\kappa
= 1$, which formally corresponds to a $1$-mode component
($\utilde{\sigma}_1$) of the symplectic spectrum for the
$2N$-by-$2N$ covariance matrix of a general Gaussian state in $N$
modes, as in the Williamson decomposition \cite{OLI12,ADE14}.

Two additional comments are deserved here: First, if a density
matrix $\hat{\rho}$ is restricted to a Gaussian form only, one can
easily relate the eigenvalues ($\utilde{\sigma}_j$) of the
covariance matrix to the eigenvalues of $\hat{\rho}$ (in the Hilbert
space) and use those eigenvalues $\utilde{\sigma}_j$ to compute
$\mbox{Tr}(\hat{\rho}^{\alpha})$ directly, without resort to our
general framework in the phase space: We now show this, for
simplicity, for the single-mode case only (note that its $N$-mode
generalization is straightforward). The most general single-mode
Gaussian state can be expressed as a displaced squeezed thermal
state (e.g., \cite{ADE14}) such that $\hat{\rho} =
\hat{D}(\alpha)\,\hat{S}(z)\,\hat{\rho}_{\beta}\,\hat{S}^{\dagger}(z)\,\hat{D}^{\dagger}(\alpha)$
where the displacement operator $\hat{D}(\alpha)$ and the squeezing
operator $\hat{S}(z)$, easily leading to
$\mbox{Tr}(\hat{\rho}^{\alpha}) =
\mbox{Tr}\{(\hat{\rho}_{\beta})^{\alpha}\}$. It is also true that
$\hat{\sigma} = \nu\,\hat{\id}$ for $\hat{\rho}_{\beta}$, where $\nu
= 2\bar{n} + 1$, and $\hat{\rho}_{\beta} = \sum_k
\{(\bar{n})^k/(\bar{n}+1)^{k+1}\}\,|k\>\<k|$, yielding
$\mbox{Tr}(\hat{\rho}^{\alpha}) = \{(\bar{n}+1)^{\alpha} -
\bar{n}^{\alpha}\}^{-1}$. This confirms the validity of Eq.
(\ref{eq:det1-2}) with ${\mathcal D}_{\sigma} = \nu^2$.

Second, it is easy to see from a harmonic oscillator in the thermal
equilibrium $\hat{\rho}_{\beta}$ [cf.
(\ref{eq:Wigner-strong-coupling-1})] that ${\mathcal D}_{\sigma} \to
4\,(\beta\hbar\omega)^{-2} + 2/3 + {\mathcal O}(\hbar^2)$ in the
limit of $\hbar \to 0$, where we used the series, $\mbox{coth}(x) =
x^{-1} + 3^{-1}\,x + {\mathcal O}(x^3)$ \cite{ABR65}. Therefore the
quantity ${\mathcal D}_{\sigma}$ and the resulting
R\'{e}nyi-$\alpha$ entropies obviously diverge within this limit, as
is the case with the von-Neumann entropy in the same limit for a
(classical) harmonic oscillator, as is well-known. This divergence
also confirms the uniform behavior of all R\'{e}nyi-$\alpha$
entropies in the classical limit tending asymptotically to the
von-Neumann entropy (also note the remark in the second last
paragraph of Sec. \ref{sec:introduction}).

Now we consider the analytic continuation of the probability moment
given in (\ref{eq:det1-2}) by extending its domain from $\{j| j =
1,2,3, \cdots\}$ to $\{\alpha| \alpha \in {\mathbb R} > 0\}$. This
enables us to introduce the R\'{e}nyi-$\alpha$ entropy of an
arbitrary $N$-mode Gaussian state such that
\begin{equation}\label{eq:det1-3}
    S_\alpha(\hat{\rho}) = \frac{\alpha\ln(2) - \ln\{F_{\sigma}(\alpha)\}}{1 -
    \alpha}
\end{equation}
with $\alpha \ne 1$, which is well-defined. Obviously, this vanishes
for all $\alpha$'s if ${\mathcal D}_{\sigma} = 1$, as required. We
are next interested in $S_1(\hat{\rho})$, which is equivalent to the
von-Neumann entropy $S_{\mbox{\sc{vN}}}(\hat{\rho})$. By applying
L'Hopital's rule to (\ref{eq:det1-3}), we can easily acquire
\begin{equation}\label{eq:det1-4}
    S_1(\hat{\rho}) = \left\{\frac{({\mathcal D}_{\sigma})^{1/2} + 1}{2}\right\}\, \ln\left\{\frac{({\mathcal D}_{\sigma})^{1/2} + 1}{2}\right\} -
    \left\{\frac{({\mathcal D}_{\sigma})^{1/2} - 1}{2}\right\}\, \ln\left\{\frac{({\mathcal D}_{\sigma})^{1/2} -
    1}{2}\right\}\,.
\end{equation}
As a special case, $S_1(\hat{{\mathsf R}}) \to S_v = (v +
1/2)\,\ln(v + 1/2) - (v -1/2)\,\ln(v - 1/2)$ for a coupled
oscillator in the thermal equilibrium state [cf.
(\ref{eq:strong-coupling-1})], being a well-known expression (e.g.,
\cite{KIM10}). In fact, it is also possible to arrive at the
expression (\ref{eq:det1-4}) independently by using Eq.
(\ref{eq:det1-5}) with (\ref{eq:det1-2}) and ${\mathcal M}_{\mu} =
\sum_{j=0}^{\mu} {\mu\choose j}\,(-\overline{p})^{\mu-j}\,
\overline{p^j}$. This accordance implies the validity of our
analytic continuation.

\section{R\'{e}nyi Entropies of $N$-Mode Non-Gaussian States}\label{sec:non-gaussian}
%
Motivated by the fact that the Gaussian state is a generalization of
the ground state $|0\>$, we now intend to discuss a generalization
of the excited states $|n\>$ where $n = 1,2,3,\cdots$. We first note
that the Wigner function of the eigenstate $|n\>$ for a single mode
is explicitly given by \cite{GRO46,BUZ95,SCH01}
\begin{equation}\label{eq:non-gaussian-2}
    W_{|n\>}(q,p) = \frac{(-1)^n}{\pi\hbar}\,\exp\left\{-\Xi(q,p)\right\}\;
    L_n\left\{2\, \Xi(q,p)\right\}\,,
\end{equation}
in which the symbol $\Xi(q,p) = \{(\hbar\kappa)^{-1}\,p\}^2 +
(\kappa q)^2$, and the Laguerre polynomial $L(x) = \sum_{k=0}^n
(-1)^k\,(k!)^{-1}\,\binom{n}{k}\,x^k$ \cite{ABR65}. Also, the Wigner
functions $W_{|n\>}$ of pure states ($n \ne 0$) are not Gaussian and
thus can be negative-valued \cite{HUD74}. If not restricted to pure
states, however, it is possible to discuss non-Gaussian mixed states
with their non-negative Wigner functions. To do so, we rewrite
(\ref{eq:non-gaussian-2}) as the linear combination
\begin{equation}\label{eq:non-gaussian-2-1}
    W_{|n\>}(q,p) = \sum_{k=0}^n a_{nk}\; W_{\underline{k}}(q,p)\,,
\end{equation}
where the coefficient $a_{nk} := (-1)^n\,(-2)^k\,\binom{n}{k}$
satisfying $\sum_{k=0}^n a_{nk} = 1$, and the Wigner function
$W_{\underline{k}}(q,p) := \Xi^k\,(k!)^{-1}\,W_{|0\>}$ (non-Gaussian
for $k \ne 0$) satisfying $\int dq\,dp\, W_{\underline{k}}(q,p) =
1$. Then for the first several values of $n$, the coefficients are
explicitly given by $a_{00}=1$; $(a_{11}=2, a_{10}=-1)$; $(a_{22}=4,
a_{21}=-4, a_{20}=1)$; $(a_{33}=8, a_{32}=-12, a_{31}=6,
a_{30}=-1)$, as well as the Wigner functions $W_{\underline{0}} =
W_{|0\>}(q,p)$; $W_{\underline{1}} = (\pi
\hbar)^{-1}\,\Xi(q,p)\,\exp\{-\Xi(q,p)\} =: W_{{\rho}_1}(q,p) \geq
0$ corresponding to the mixed state $\hat{\rho}_1 = (|0\>\<0| +
|1\>\<1|)/2$; $W_{\underline{2}} =
(2\,\pi\hbar)^{-1}\,\{\Xi(q,p)\}^2\,\exp\{-\Xi(q,p)\} =:
W_{\rho_2}(q,p) \geq 0$ corresponding to the mixed state
$\hat{\rho}_2 = (|0\>\<0| + 2\,|1\>\<1| + |2\>\<2|)/4$.

Therefore we next consider, as a generalization of the single-mode
Wigner functions $W_{\underline{k}}(q,p)$, the $N$-mode non-Gaussian
states in form of
\begin{equation}\label{eq:non-gaussian-state1-1}
    W_{\rho_{\nu}}(\vec{x}) = \frac{\exp\{-(\vec{x})^{\scriptscriptstyle T}\,(\hat{\sigma})^{-1}\,
    \vec{x}\}}{{(N)_{\nu}\;\pi^{\scriptscriptstyle
    N}\,\{\det(\hat{\sigma})\}}^{1/2}}\, \{(\vec{x})^{\scriptscriptstyle T}\,(\hat{\sigma})^{-1}\,\vec{x}\}^{\nu}\,,
\end{equation}
where $\nu = 1,2,3,\cdots$, and the Pochhammer symbol $(N)_{\nu} =
\Gamma(N+\nu)/\Gamma(N)$ \cite{ABR65} (note that an $N$-mode
generalization of $W_{|n\>}(q,p)$ will be discussed below in terms
of the linear combination of $W_{\rho_{\nu}}(\vec{x})$'s). This
Wigner function (\ref{eq:non-gaussian-state1-1}) also is fully
determined by the matrix elements $\sigma_{jk}$. For the normalizing
(i.e., $\int d^{2{\scriptscriptstyle N}}\vec{x}\;
W_{\rho_{\nu}}(\vec{x}) \stackrel{!}{=} 1$), we here applied the
Gaussian integral (\ref{eq:gaussian-integral_zee}) with the help of
$e^{-y^2} y^{2\nu} = (-\partial_a)^{\nu}\,e^{-a y^2}|_{a=1}$ and
$(-\partial_a)^{\nu}\,a^{-{\scriptscriptstyle N}}|_{a=1} =
(N)_{\nu}$. To acquire the first moment of probability
$\overline{p}_{\nu} = \mbox{Tr}(\hat{\rho}_{\nu}^2)$ in closed form,
we now discuss the integral
\begin{equation}\label{eq:non-gaussian-integral-1}
    I_{\mu\nu}\, =\, (2\pi\hbar)^{\scriptscriptstyle N} \int d^{2{\scriptscriptstyle N}}\vec{x}\; W_{\rho_{\mu}}(\vec{x})\;
    W_{\rho_{\nu}}(\vec{x})\, =\, \frac{1}{C_{\mu\nu}\, ({\mathcal
    D}_{\sigma})^{1/2}}\,,
\end{equation}
in which $C_{\mu\nu} = 2^{\mu+\nu}\,(N)_{\nu}/(N+\mu)_{\nu} =
C_{\nu\mu}$. It then follows that $\overline{p}_{\nu} = I_{\nu\nu}$
and $S_2(\hat{\rho}_{\nu}) = 2^{-1}\,\ln({\mathcal D}_{\sigma,\nu})$
where ${\mathcal D}_{\sigma,\nu} = (C_{\nu\nu})^2\, {\mathcal
D}_{\sigma} =
\{\mbox{det}(\hat{\sigma}_{\nu})\}/h^{2{\scriptscriptstyle N}}$
expressed in terms of the matrix $\hat{\sigma}_{\nu} :=
(C_{\nu\nu})^{1/{\scriptscriptstyle N}}\,\hat{\sigma}$. For a later
purpose, we briefly take into consideration the simplest case of
$N=1$. Then we have the entropy in reduced form,
$S_2(\hat{\rho}_{\nu}) \to 2^{-1}\,\ln({\mathcal D}_{\sigma}) +
\ln\{\pi^{1/2}\,\Gamma(\nu+1)/\Gamma(\nu+1/2)\}$, where we applied
the identity, $\Gamma(2x) =
(2\pi)^{-1/2}\,2^{2x-1/2}\,\Gamma(x)\,\Gamma(x+1/2)$ \cite{ABR65}.
As seen, this reduced form is positive-valued so long as $\nu \geq
1$, showing that $\hat{\rho}_{\nu}$ denotes a mixed state, even for
the case of ${\mathcal D}_{\sigma} = 1$, by construction. Figs.
\ref{fig:fig3} and \ref{fig:fig4} plot the behaviors of
$S_2(\hat{\rho}_{\nu})$ versus $\nu$ and $N$, respectively.

Interestingly, we recognize here that the above result, valid for
the non-Gaussian state (\ref{eq:non-gaussian-state1-1}), can exactly
be recovered by introducing an effective Gaussian state (of that
non-Gaussian state) with its covariance matrix
$\utilde{\hat{\sigma}}_{\nu} = 2^{-1}\,\hat{\sigma}_{\nu}$, for
which the Wigner entropy is obviously well-defined and explicitly
given by $S_{\scriptscriptstyle W}(\hat{\rho}_{\mbox{\sc eff}}) = N
+ N \ln(\pi) + 2^{-1} \ln({\mathcal D}_{\sigma,\nu})$, thus
coinciding with $S_2(\hat{\rho}_{\nu})$ up to a constant again. This
coincidence, valid also for the non-Gaussian states in an extended
sense, may be regarded as a generalization of the same coincidence,
as is well-known, valid for the Gaussian states.

Next we consider an arbitrary linear combination of the states given
in (\ref{eq:non-gaussian-state1-1}) such that $W_{\rho}(\vec{x}) =
\sum_{\nu} a_{\nu}\, W_{\rho_\nu}(\vec{x})$ where $a_{\nu} \in
{\mathbb R}$ and $\sum a_{\nu} = 1$ but $a_{\nu}$'s are not
necessarily non-negative and so the resulting Wigner function
$W_{\rho}(\vec{x})$ can be negative-valued. Applying
(\ref{eq:non-gaussian-integral-1}), it is straightforward to find
that $S_2(\hat{\rho}) = 2^{-1} \ln(\overline{\mathcal D}_{\sigma})$
where $\overline{\mathcal D}_{\sigma} := (\sum_{\mu,\nu}
a_{\mu}\,a_{\nu}/C_{\mu\nu})^{-2}\,{\mathcal D}_{\sigma}$. Likewise,
the Wigner entropy of its effective Gaussian state with
$\hat{\sigma} \to (\sum_{\mu,\nu}
a_{\mu}\,a_{\nu}/C_{\mu\nu})^{-1/{\scriptscriptstyle
N}}\,\hat{\sigma} =: \hat{\underline{\sigma}}$ is exactly given by
the expression of $S_{\scriptscriptstyle W}(\hat{\rho}_{\mbox{\sc
eff}})$ in the preceding paragraph but simply with ${\mathcal
D}_{\sigma,\nu} \to \overline{\mathcal D}_{\sigma}$.

Now we restrict our discussion into the particular case that the
coefficients $a_{\nu} \to a_{nk}$ given in
(\ref{eq:non-gaussian-2-1}) for a given value of $n$. Then the
resulting Wigner function $W_{\rho}(\vec{x}) = \sum_k a_{nk}\,
W_{\rho_k}(\vec{x})$ can be regarded, by construction, as a
generalization of $W_{|n\>}(q,p)$ in (\ref{eq:non-gaussian-2-1}),
i.e., $W_{\underline{k}}(q,p)$ for a single mode replaced by
$W_{\rho_k}(\vec{x})$ in (\ref{eq:non-gaussian-state1-1}) for $N$
modes; as a simple example, $W_{\rho}(\vec{x}) =
a_{11}\,W_{\rho_0}(\vec{x}) + a_{10}\,W_{\rho_1}(\vec{x})$ for
$n=1$, which is a generalization of $W_{|1\>}(q,p)$. The explicit
expressions of $a_{nk}$ and $C_{\mu\nu}$ will give, after some
algebraic manipulations, the reduced expression, $(\sum_{\mu,\nu}
a_{\mu}\,a_{\nu}/C_{\mu\nu})^{-1} \to \binom{n+N-1}{N-1}$. This
immediately yields the entropy $S_2(\hat{\rho}) = 2^{-1}
\ln{\mathcal D}_{\sigma} + \ln\binom{n+N-1}{N-1}$ in reduced form,
in which the second term increases with $n$ and $N$. It is also
instructive to note that for the case of $N=1$, the quantity
$\binom{n+N-1}{N-1}$ reduces to unity, regardless of the value of
$n$, therefore the entropy $S_2(\hat{\rho}) \to 2^{-1} \ln({\mathcal
D}_{\sigma})$. This entropy will vanish not only for the pure
Gaussian state $|0\>$ with ${\mathcal D}_{\sigma} = 1$, but also,
remarkably, for the pure non-Gaussian state $|n\>$ with $n \ne 0$,
as required, due to the fact that in this case, by construction, the
effective covariance matrix $2^{-1}\,\hat{\underline{\sigma}}$
reduces to its Gaussian-state counterpart $2^{-1}\,\hat{\sigma}$
indeed, independently of $n$! This shows a robust structure of the
current generalization into the non-Gaussian states. It may also be
legitimate to say that this generalization of the basis states
$|n\>$'s will shed new light on the study of the R\'{e}nyi entropies
for a broader class of non-Gaussian states to be studied later.

Next let us briefly discuss higher-order moments $\overline{p^j}$
with $j = 2,3,4,\cdots$, and the resulting entropies
$S_{j+1}(\hat{\rho})$ for the non-Gaussian states. To apply the same
techniques including the Gaussian integral
(\ref{eq:gaussian-integral_zee}) as for $S_2(\hat{\rho})$, we
express the Wigner function as
\begin{equation}\label{eq:ggaussian-state1-1}
    W_{\rho}(\vec{x}) = \frac{\exp\{- a\,(\vec{x})^{\scriptscriptstyle T} (\hat{\sigma})^{-1}\,\vec{x}\}}{{\pi^{\scriptscriptstyle N}
    \{\det(\hat{\sigma})\}}^{1/2}}
\end{equation}
up to the normalizing, where the parameter $a$ will be set unity.
Then, the Fourier transform is in form of
\begin{equation}\label{eq:ggaussian-state1-1-1}
    \widetilde{W}_{\rho}(\vec{x}) =
    \frac{\exp\{-(4a\hbar^2)^{-1}\,(\vec{x})^{\scriptscriptstyle T}
    \hat{\Lambda}\,\hat{\sigma}\,\hat{\Lambda}\,\vec{x}\}}{(2 \pi \hbar\,a)^{\scriptscriptstyle
    N}}\,.
\end{equation}
Using (\ref{eq:ggaussian-state1-1}) and
(\ref{eq:ggaussian-state1-1-1}) in place of
(\ref{eq:gaussian-state1-1}) and (\ref{eq:gaussian-state1-1-1}), it
is straightforward to acquire
\begin{equation}\label{eq:jth-moment-1-1}
    \overline{p^j} = (2\hbar)^{-j{\scriptscriptstyle N}}\, ({\mathcal D}_\sigma)^{-1/2}\,
    \{\mbox{det}(\hat{D}_{j+1})\}^{-1/2}\, (a_1)^{-\scriptscriptstyle N}\,(a_2)^{-\scriptscriptstyle N} \cdots
    (a_j)^{-\scriptscriptstyle N}
\end{equation}
[cf. (\ref{eq:jth-moment-1})] where $\mbox{det}(\hat{D}_{j+1})$
given in Tab. \ref{tab:table1} but simply with the replacement of
$\hat{\sigma} \to (a_{\mu})^{-1}\,\hat{\sigma}$ where $\mu = 1,2,
\cdots j$, as well as with $(\hat{\sigma})^{-1} \to
a\,(\hat{\sigma})^{-1}$ in its diagonal elements. Eq.
(\ref{eq:jth-moment-1-1}) can be evaluated exactly the same way
(\ref{eq:jth-moment-1}) was evaluated in Sec. \ref{sec:gaussian}.
Then we perform the differentiation
$(-\partial_a)^{\nu}\,(-\partial_{a_1})^{\nu}\,(-\partial_{a_2})^{\nu}
\cdots (-\partial_{a_j})^{\nu}$ and set $a, a_{\mu} = 1$, followed
by inserting the normalizing constants $\{(N)_{\nu}\}^{-1}$'s. Its
explicit expression and that of the resulting R\'{e}nyi entropy are
straightforward to evaluate but simply too large in size, and so
here, e.g., only for $N=1$ and $\nu = 1,2,3$,
\begin{eqnarray}\label{eq:examples}
    \mbox{Tr}\{(\hat{\rho}_1)^3\} &=& \frac{4}{3^2\,z^4}\,(2 z^3 + 3 z^2 - 12 z + 16)\n\\
    \mbox{Tr}\{(\hat{\rho}_2)^3\} &=&
    \frac{4}{3^4\,z^7}\,(10 z^6 + 12 z^5 - 3 z^4 - 74 z^3 + 456 z^2 - 960 z + 640)\n\\
    \mbox{Tr}\{(\hat{\rho}_3)^3\} &=&
    \frac{4}{3^8\,z^{10}}\,(560 z^9 + 630 z^8 + 180 z^7 - 1221 z^6 + 252 z^5 + 30960 z^4 - 190560 z^3\n\\
    && +\,524160 z^2 -
    645120 z + 286720)\,,
\end{eqnarray}
and so on, where $z := 3{\mathcal D}_{\sigma} + 1$ (cf. Fig.
\ref{fig:fig5} for $N = 1$ and Fig. \ref{fig:fig6} for $N = 2$). For
comparison, $\mbox{Tr}\{(\hat{\rho}_0)^3\} = 4/z$ (cf. Tab.
\ref{tab:table2}).

Finally, we study the probability moments expressed in terms of the
Husimi function briefly. In fact, it is highly tempting to do so,
especially for non-Gaussian states, due to its inherent non-negative
feature. For simplicity, we primarily restrict our discussion to the
case of a single mode. The Husimi function can be understood as the
Weierstrauss transform of the Wigner function, i.e., the convolution
of the Wigner function with a Gaussian filter such that \cite{LEE95}
\begin{eqnarray}\label{eq:husimi_2}
    Q_{\rho}(q,p) &=& \frac{1}{\pi\hbar} \int dq' dp'\;
    W_{\rho}(q',p')\, \exp\left\{-\left(\{\kappa\,(q'-q)\}^2 +
    \left(\frac{p'-p}{\hbar\kappa}\right)^2\right)\right\}\,,
\end{eqnarray}
as well as
\begin{eqnarray}\label{eq:identity_phase_1}
    W_{\rho}(q,p) &=& \exp\{-(a\,\partial_q^2 + b\,\partial_p^2)\}\,
    Q_{\rho}(q,p)\n\\
    &=& \int_{-\infty}^{\infty} \frac{dq' dp'}{2\pi\hbar}\,
    \exp\left\{\frac{a (p')^2}{\hbar^2} + i \frac{p' q}{\hbar}\right\}
    \exp\left\{\frac{b (q')^2}{\hbar^2} + i \frac{q' p}{\hbar}\right\}\,\widetilde{Q}_{\rho}(q',p')
\end{eqnarray}
where $a = (4 \kappa^2)^{-1}$ and $b = (\hbar\kappa)^2/4$, and the
Fourier transform
\begin{equation}\label{eq;husimi-2}
    \widetilde{Q}_{\rho}(q',p') = \int \frac{dq dp}{2\pi\hbar}\; Q_{\rho}(q,p)\,
    \exp\left\{\frac{-i (q p' + p q')}{\hbar}\right\}\,.
\end{equation}
Then a useful identity will follow,
\begin{equation}\label{eq:wigner-husimi-1}
    \widetilde{W}_{\rho}(q,p) =
    \exp\left(\frac{a p^2 + b x^2}{\hbar^2}\right)\;
    (\widetilde{Q}_{\rho})^{\ast}(q,p)\,,
\end{equation}
where the complex conjugate $(\widetilde{Q}_{\rho})^{\ast}(q,p) =
\widetilde{Q}_{\rho}(-q,-p)$. Substituting
(\ref{eq:identity_phase_1}) and (\ref{eq:wigner-husimi-1}) into
(\ref{eq:bopp_4-3-1}), we can obtain
\begin{eqnarray}
    \overline{p} &=& (2\pi\hbar) \int dq' dp'\,
    e^{\{2a\,(p')^2 + 2b\,(q')^2\}/\hbar^2}\; |(\widetilde{Q}_{\rho})^{\ast}(q',p')|^2\label{eq:renyi-husimi-0}\\
    \overline{p^2} &=& (2\pi\hbar) \int dq_1\,dp_1\,dq_2\,dp_2\;
    e^{(2a\,\{(p_1)^2+(p_2)^2 + p_1\,p_2\}\,+\,2b\,\{(q_1)^2+(q_2)^2 +
    q_1\,q_2\})/\hbar^2} \times\n\\
    && \widetilde{Q}_{\rho}(q_1+q_2, p_1+p_2)\cdot (\widetilde{Q}_{\rho})^{\ast}(q_1,p_1)\cdot
    (\widetilde{Q}_{\rho})^{\ast}(q_2,p_2)\cdot
    \exp\left\{\frac{i (q_1 p_2 - p_1 q_2)}{2\hbar}\right\}\,,\label{eq:renyi-husimi-0-1}
\end{eqnarray}
and so on.

Several additional comments are appropriate here. First, from the
single-mode results such as (\ref{eq:renyi-husimi-0}) and
(\ref{eq:renyi-husimi-0-1}), the $N$-mode counterparts will follow
straightforwardly. Second, it is easy to show, with the help of
(\ref{eq:husimi_2}), that if the Wigner function is Gaussian, then
the Husimi function is Gaussian, too. Third, while the von-Neumann
entropy of any pure state is zero, this is not the case for the
Wehrl entropy defined as $(-1) \int dq dp\; Q_{\rho}(q,p)\,
\ln\{Q_{\rho}(q,p)\}$ (for a non-Gaussian state) in terms of the
Husimi function $Q_{\rho} = \<\beta|\hat{\rho}|\beta\>/\pi$ where
$\beta = 2^{-1/2}\,(\kappa q + i p/\hbar\kappa)$, fundamentally due
to the non-orthogonality $|\<\beta|\gamma\>|^2 =
e^{-|\beta-\gamma|^2}$ \cite{GNU01}. Exploring the R\'{e}nyi
entropies in terms of the Wehrl entropy for the wild world of
non-Gaussian states remains an open question.

\section{Concluding Remarks}\label{sec:conclusion}
%
We have studied the R\'{e}nyi-$\alpha$ entropies
$S_{\alpha}(\hat{\rho})$ for both Gaussian and non-Gaussian states
in $N$ modes. We have first derived the entropies of Gaussian states
in closed form for arbitrary positive integers $j \leftarrow \alpha$
and then acquired them also for real values of $\alpha
> 0$ with the help of a recurrence relation between those entropies
for two consecutive values of $j$'s and then the analytic
continuation. In the literature, on the other hand, the entropy
$S_2(\hat{\rho})$ has been the primary quantity thus far. In fact,
this entropy $S_2(\hat{\rho})$ results from the first-order moment
of probability $\overline{p}$, as shown, and accordingly contains a
``coarse-grained'' information only. Although the statistical
behaviors of Gaussian states are, as is well-known, determined
essentially by its covariance matrix alone, thus giving rise to
$S_2(\hat{\rho})$ directly, the concrete shapes of higher-order
moments $\overline{p^{\alpha}} = \mbox{Tr}(\hat{\rho}^{\alpha+1})$
have been unknown therefore have not yet been intensively explored.

Subsequently, we have studied the R\'{e}nyi-$\alpha$ entropies for
an interesting class of non-Gaussian states which can be regarded as
a generalization of the eigenstates of a single harmonic oscillator.
By introducing the effective Gaussian states for the non-Gaussian
states, we have also generalized the same relation between the
first-moment entropy $S_2(\hat{\rho})$ and the Wigner entropy of its
(effective) Gaussian state as other investigators have shown
previously only for the case of Gaussian states. Because the
dynamics of an $N$-oscillator system is Gaussian, our result will
contribute to a systematic study of the entropy dynamics when the
current form of a non-Gaussian state is initially prepared. The next
subject will include an investigation of entanglement in terms of
the R\'{e}nyi entropies, also for a wider class of non-Gaussian
states, as well as a systematic analysis of the quantum-classical
transition for those entropies in the semiclassical limit, which
could also provide an interesting and novel contribution to the
study of open systems.

\section*{Acknowledgments}
The author appreciates all comments and constructive questions of
the referees which made him clarify the paper and improve its
quality, especially in relation to the two comments after Eq.
(\ref{eq:Wigner-strong-coupling-1}). He gratefully acknowledges the
financial support provided by the U.S. Army Research Office (Grant
No. W911NF-15-1-0145).

\appendix*\section{Derivation of the generator in closed form: Eq. (\ref{eq:det1-1})}\label{sec:appendix}
%
We explicitly evaluate the generator $\mbox{det}(\hat{G}_j) =
\mbox{det}\{\hat{A} - \hat{B}_j\,(\hat{D}_j)^{-1}\,\hat{C}_j\}$,
starting from the case of $j=1$, followed by $j = 2, 3, 4, \cdots$,
first for a single mode, followed by multi modes (cf. Tab.
{\ref{tab:table1} and Fig. \ref{fig:fig1}). We can then find the
recurrence relation
\begin{equation}\label{eq:appendix-1}
    \mbox{det}(\hat{G}_j) = \frac{{\mathcal
    D}_{\sigma}}{(2\hbar)^{2{\scriptscriptstyle N}}}\,\left\{\frac{1 + U(j)}{2}\right\}^2
\end{equation}
where ${\mathcal D}_{\sigma} =
(\mbox{det}\,\hat{\sigma})/\hbar^{2{\scriptscriptstyle N}}$ and
\begin{equation}\label{eq:appendix-2}
    U(j) = 1 + \frac{({\mathcal D}_{\sigma})^{-1} - 1}{U(j-1) + 1}\,.
\end{equation}
Consider $U(j) + c = \lambda\,\{U(j-1) + b\}$, where $\lambda = (1 +
c)/\{U(j-1) + 1\}$, and $b = \{c + ({\mathcal D}_{\sigma})^{-1}\}/(1
+ c)$. Requiring now that $ b \stackrel{!}{=} c$. Then it follows
that $c = ({\mathcal D}_{\sigma})^{-1/2} =: c_0$. Let $Y(j) = \{U(j)
+ c_0\}^{-1}$. This satisfies the recurrence relation
\begin{equation}\label{eq:appendix-3}
    Y(j) = \alpha\, Y(j-1) + \beta\,,
\end{equation}
where $\alpha = \{({\mathcal D}_{\sigma})^{1/2} - 1\}/\{({\mathcal
D}_{\sigma})^{1/2} + 1\}$, and $\beta = ({\mathcal
D}_{\sigma})^{1/2}/\{({\mathcal D}_{\sigma})^{1/2} + 1\}$. This
means that $Y(j) - \beta/(1-\alpha)$ is a geometric sequence with
the common ratio $r = \alpha$, which will result in
\begin{equation}\label{eq:appendix-4}
    Y(j) = \frac{({\mathcal D}_{\sigma})^{1/2}}{2} \left\{1 -
    \left(\frac{{(\mathcal D}_{\sigma})^{1/2} - 1}{({\mathcal D}_{\sigma})^{1/2} +
    1}\right)^j\right\}\,.
\end{equation}
Consequently, we can determine the closed form of $U(j)$ and then
that of $\mbox{det}(\hat{G}_j)$ in (\ref{eq:appendix-1}) and
(\ref{eq:det1-1}).
\newpage
\begin{figure}[htb]
\centering\hspace*{-0cm}\vspace*{-1cm}{
\includegraphics[scale=0.7]{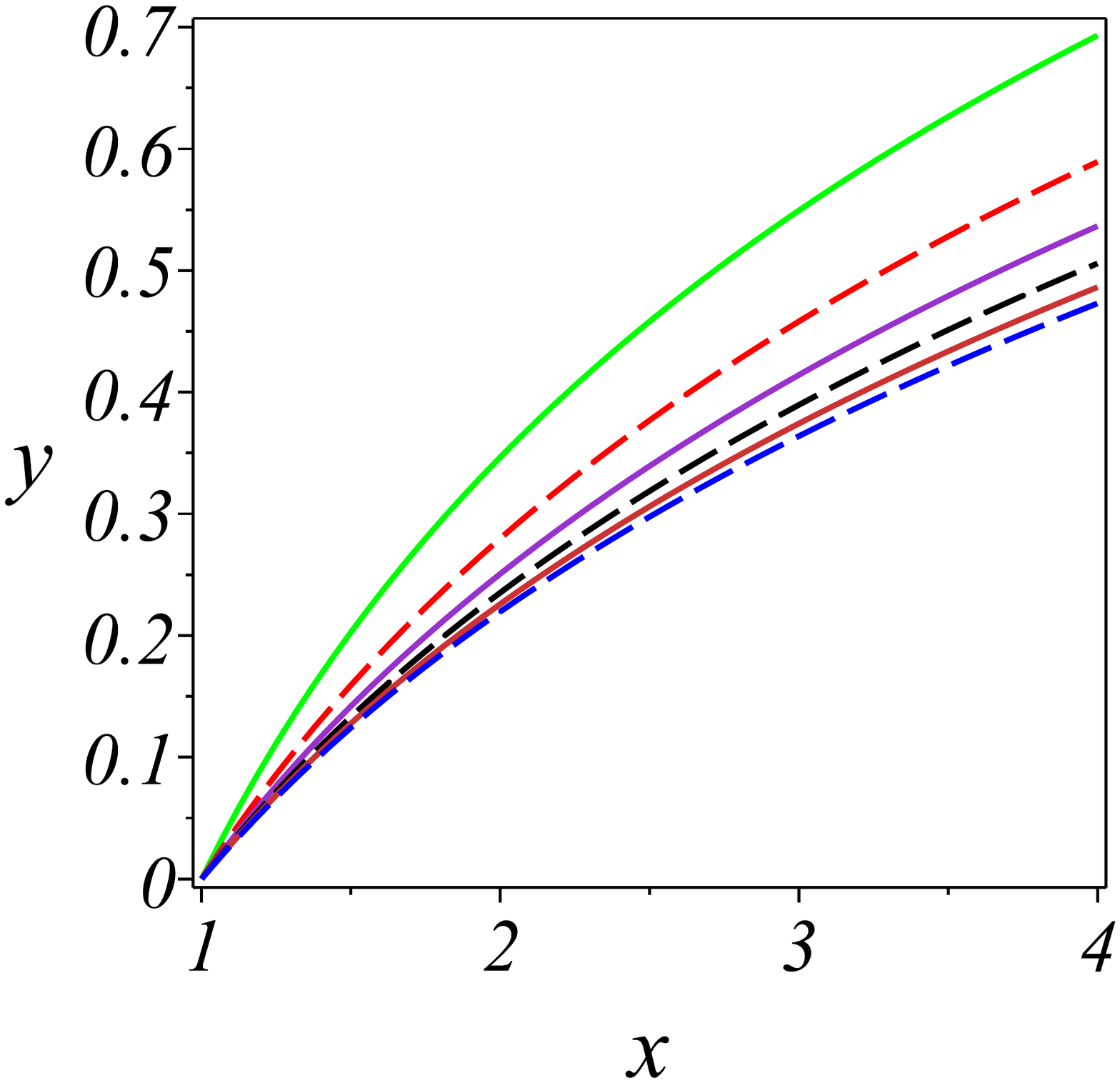}
\caption{\label{fig:fig2}}}
\end{figure}
Fig.~\ref{fig:fig2}: (Color online) The R\'{e}nyi-$\alpha$ entropies
for $N$ modes in Tab. \ref{tab:table2}, $y = S_{\alpha}(\hat{\rho})$
versus $x = {\mathcal D}_{\sigma}$ for $\hat{\rho}$. The values
$\alpha = 2,3,4,5,6,7$, in sequence from top to bottom. As required,
$y=0$ at $x=1$ for all $\alpha$'s.
\newpage
\begin{figure}[htb]
\centering\hspace*{-0cm}\vspace*{-0cm}{
\includegraphics[scale=0.7]{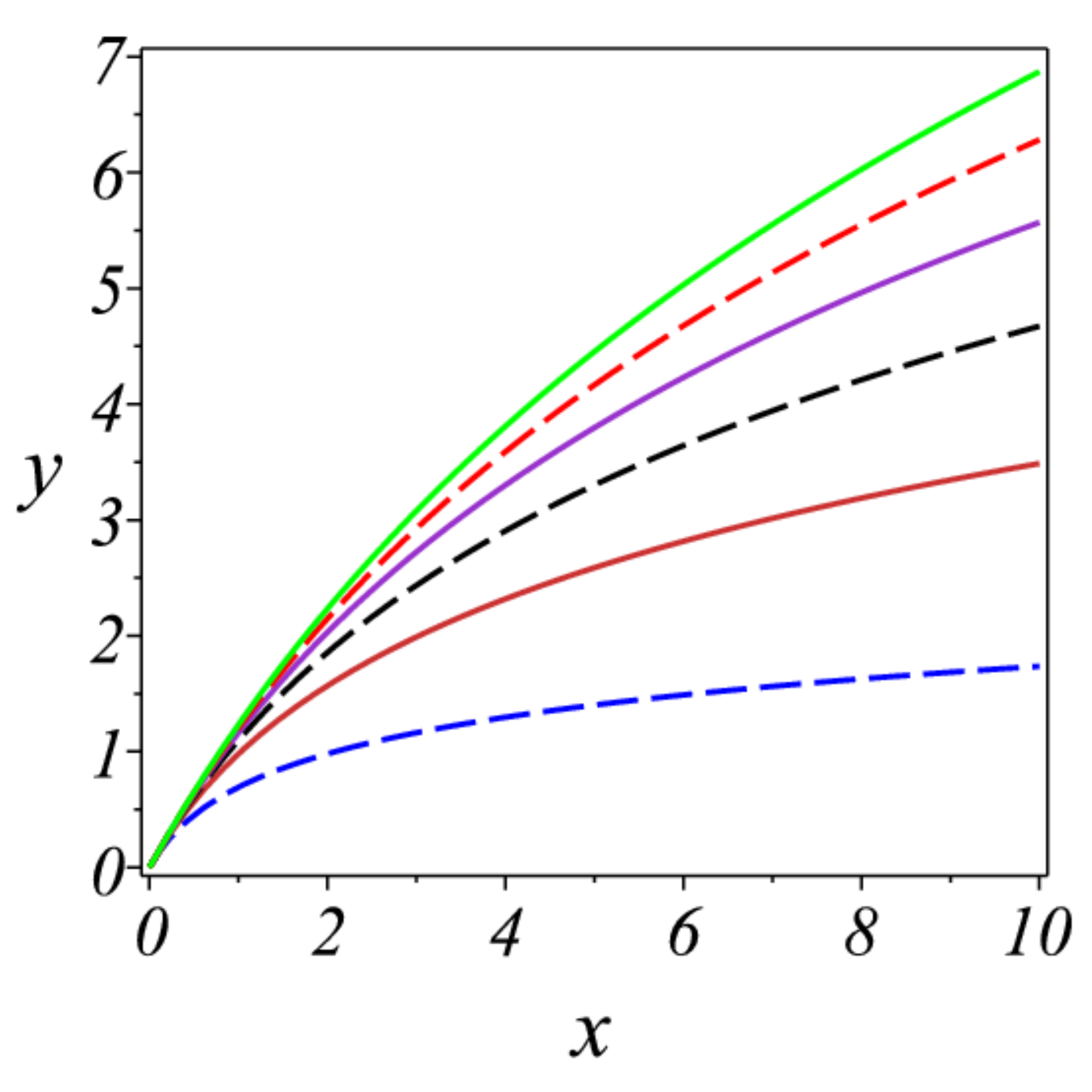}
\caption{\label{fig:fig3}}}
\end{figure}
Fig.~\ref{fig:fig3}: (Color online) $y = S_2(\hat{\rho}_{\nu}) -
S_2(\hat{\rho}_0) = \ln\{4^{\nu}\,(N)_{\nu}/(N+\nu)_{\nu}\}$ versus
$x = \nu$, in discussion after Eq.
(\ref{eq:non-gaussian-integral-1}). Here the state $\hat{\rho}_0$
denotes a Gaussian state with $S_2(\hat{\rho}_0) = 2^{-1}
\ln({\mathcal D}_{\sigma})$, and the substitution of $x$ for integer
$\nu$ may be considered the analytic continuation. The values $N =
1,2,3,4,5,6$, in sequence from bottom to top. For a given value of
${\mathcal D}_{\sigma}$, as seen, each curve increases with $x$;
especially if ${\mathcal D}_{\sigma} = 1$ (denoting a pure Gaussian
state), then the non-Gaussian state $\hat{\rho}_{\nu}$ with $\nu
> 0$ and $W_{\rho_{\nu}}(\vec{x}) \geq 0$ [cf.
(\ref{eq:non-gaussian-state1-1})] denotes a non-pure state by
construction, so its entropies should be larger than those of the
Gaussian counterparts. Note an explicit $N$-dependence of $y$ here,
while it is not the case for Gaussian states ($\nu = 0$).
\newpage
\begin{figure}[htb]
\centering\hspace*{-0cm}\vspace*{-0cm}{
\includegraphics[scale=0.7]{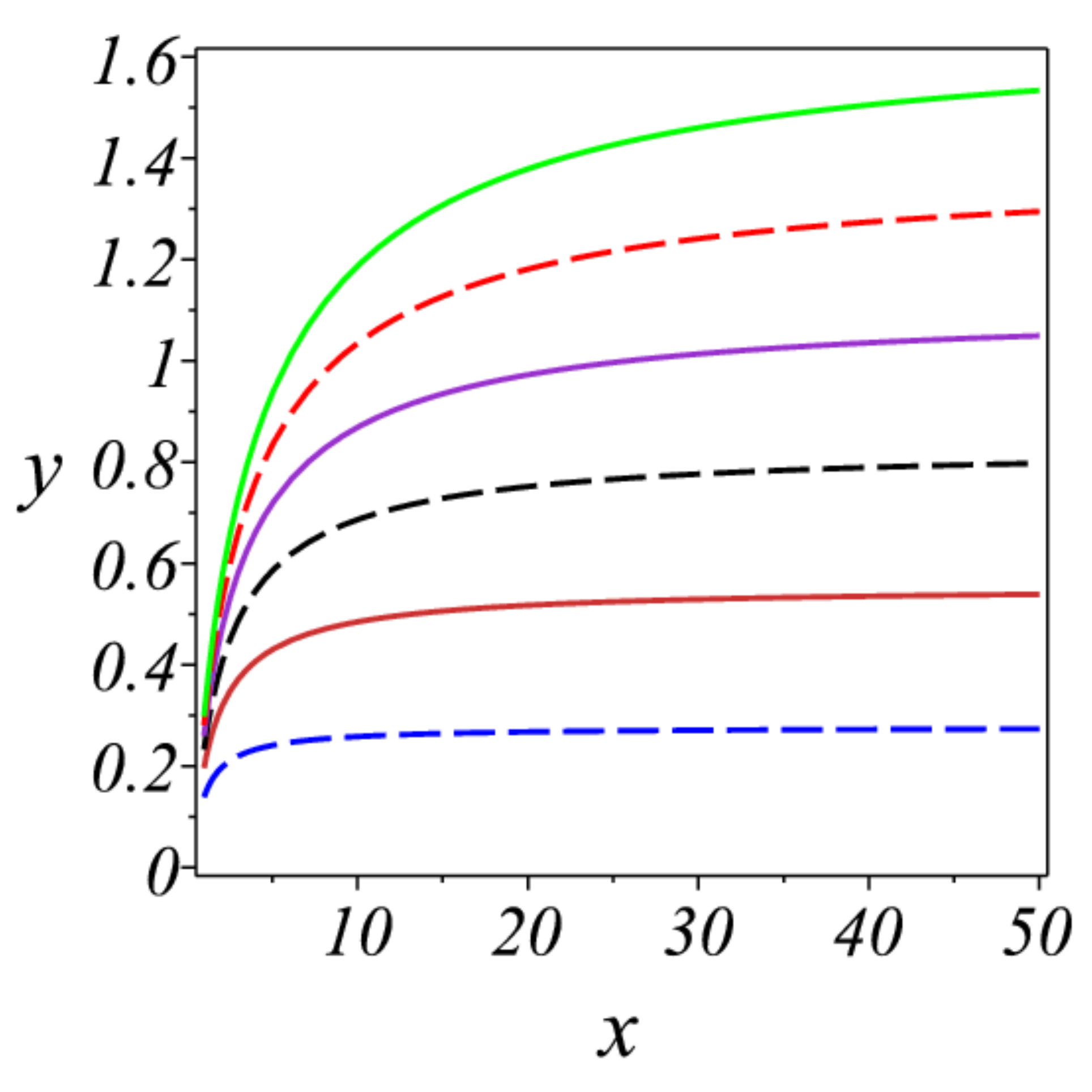}
\caption{\label{fig:fig4}}}
\end{figure}
Fig.~\ref{fig:fig4}: (Color online) $y =
0.2\,\{S_2(\hat{\rho}_{\nu}) - S_2(\hat{\rho}_0)\} =
0.2\,\ln\{4^{\nu}\,(N)_{\nu}/(N+\nu)_{\nu}\}$ (re-scaled) versus $x
= N$, as for Fig. \ref{fig:fig3}. The values $\nu = 1,2,3,4,5,6$, in
sequence from bottom to top. As seen, $y \to 0.2\,\ln(4^{\nu})$ with
$N \to \infty$.
\newpage
\begin{figure}[htb]
\centering\hspace*{-3cm}\vspace*{-0cm}{
\includegraphics[scale=0.7]{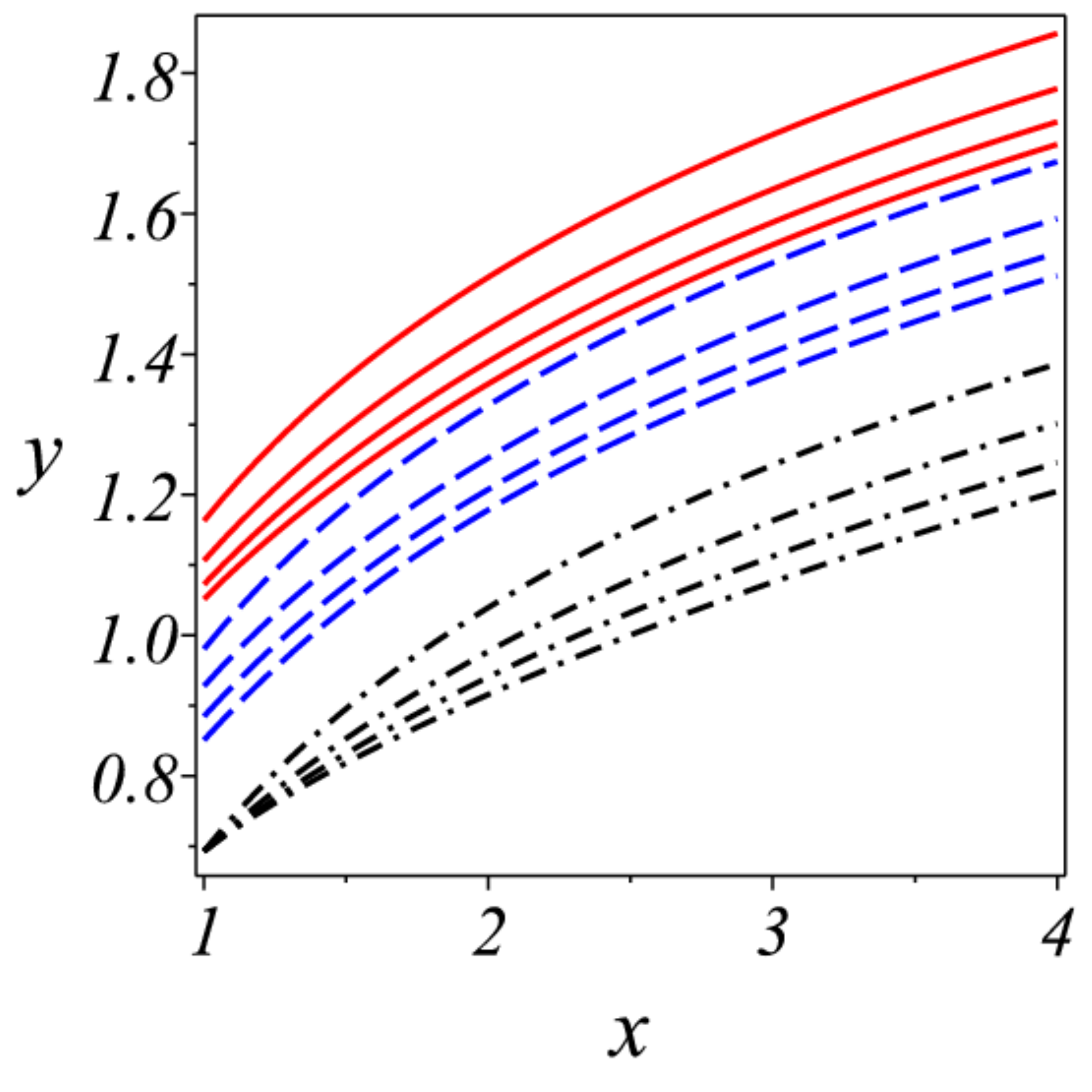}
\caption{\label{fig:fig5}}}
\end{figure}
Fig.~\ref{fig:fig5}: (Color online) The R\'{e}nyi entropies $y =
S_{\alpha}(\hat{\rho}_{\nu})$ versus the $x = {\mathcal D}_{\sigma}$
for the non-Gaussian states with $N = 1$ [cf. Eqs.
(\ref{eq:jth-moment-1-1}) and (\ref{eq:examples})]. From top to
bottom, 1) the solid curves with $\nu = 3$: $\alpha = 2,3,4,5$, in
sequence from top to bottom; 2) the dash curves with $\nu = 2$, in
the same way; 3) the dashdot curves with $\nu = 1$, in the same way.
\newpage
\begin{figure}[htb]
\centering\hspace*{-3cm}\vspace*{-0cm}{
\includegraphics[scale=0.7]{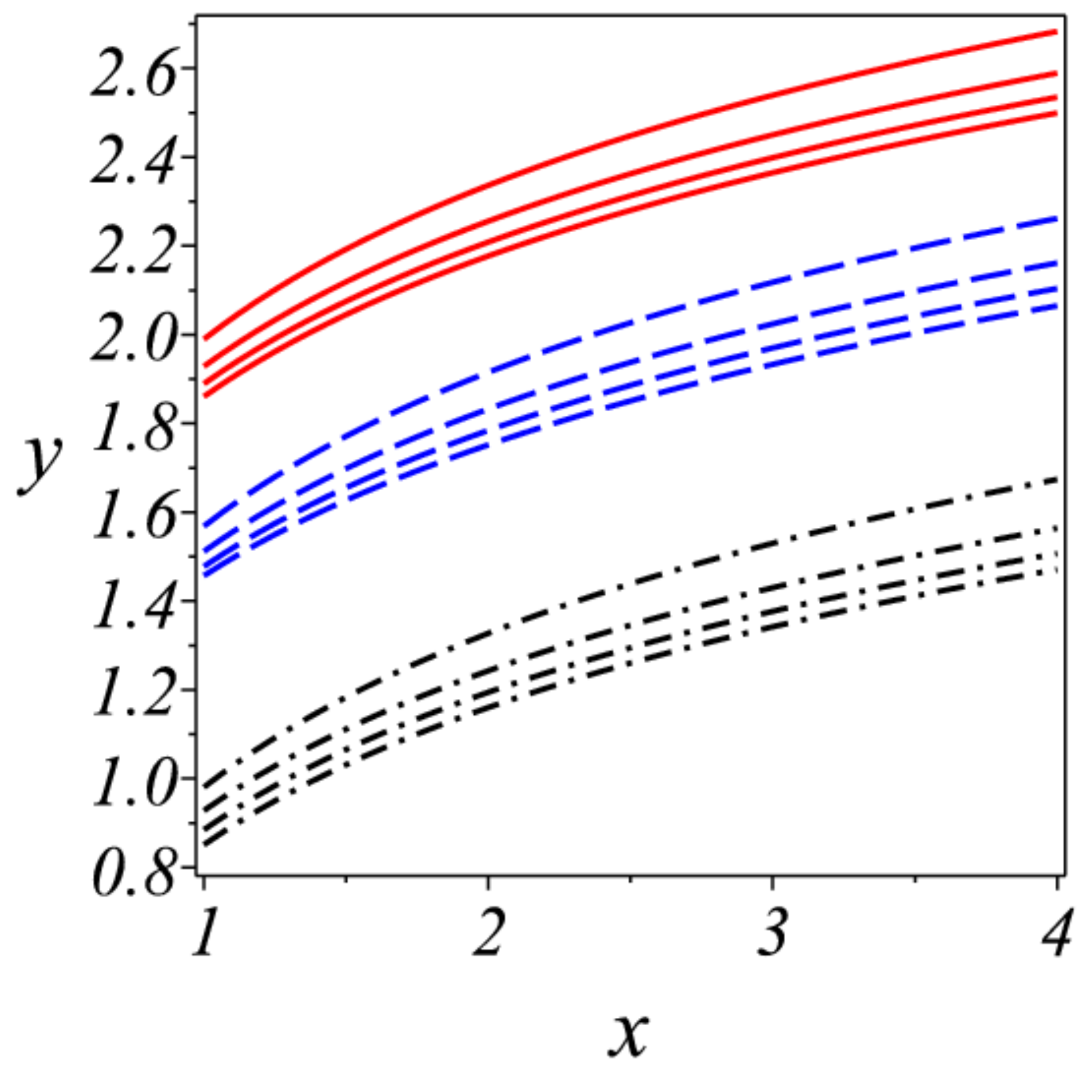}
\caption{\label{fig:fig6}}}
\end{figure}
Fig.~\ref{fig:fig6}: (Color online) (Color online) The R\'{e}nyi
entropies $y = S_{\alpha}(\hat{\rho}_{\nu})$ versus the $x =
{\mathcal D}_{\sigma}$ for $N = 2$. Otherwise, the same parameters
as for Fig. \ref{fig:fig5}.
\end{document}